# Fermat Number Transform Based Chromatic Dispersion Compensation and Adaptive Equalization Algorithm

Siyu Chen[(1)], Zheli Liu[(1)], Weihao Li[(1)], Zihe Hu[(1)], Mingming Zhang[(1)], Sheng Cui[(1)] and Ming Tang[(1)]

[(1)] Wuhan National Lab for Optoelectronics (WNLO) & National Engineering Research Centre of Next Generation Internet Access-system, School of Optical and Electronic Information, Huazhong University of Science and Technology, Wuhan, 430074, China, tangming@mail.hust.edu.cn

**Abstract** *By introducing the Fermat number transform into chromatic dispersion compensation and adaptive equalization, the computational complexity has been reduced by 68% compared with the conventional implementation. Experimental results validate its transmission performance with only 0.8 dB receiver sensitivity penalty in a 75 km-40 GBaud-PDM-16QAM system. ©2024 The Author(s)*

**Introduction**

With the vigorous development of artificial intelligence technology in recent years, coherent optical communication, due to its high spectral efficiency and capability to compensate various transmission impairments in intra and inter-data centre scenarios, is considered a promising direction for optical communication in future data centres [1].

However, high complexity and power consumption limit the widespread application of coherent optical communication in the context of data centre inter-connection. In existing coherent modules, the power consumption of the digital signal processing (DSP) accounts for approximately 50% of the total power consumption [2]. Among DSP algorithms at the receiver, the chromatic dispersion compensation (CDC) and adaptive equalization (AEQ) consume a significant amount of resource. When the length of filter taps exceeds a threshold of 10, the frequency domain convolution is more efficient than the time domain convolution [3]. Thus, frequency domain adaptive equalization (FD-AEQ) and frequency domain chromatic dispersion compensation (FD-CDC) are more efficient than time domain adaptive equalization (TD-AEQ) and time domain chromatic dispersion compensation (TD-CDC) in systems with larger communication capacities. FD-CDC&AEQ transform the signal into the frequency domain via FFT first, multiply it by tap coefficients, and then transform it back into the time domain through IFFT. However, when performing the FFT and IFFT on a sequence of length $N$, it still requires $N\log_2 N$ multiplication operations, which consumes a large amount of resource. The Fermat number transform (FNT), as a promising method to further simplify the complexity, has been introduced into CDC through simulation verification [4]. However, the method is far from mature to be applied in data centre interconnections.

In this paper, we proposed to apply FNT into CDC and AEQ algorithms. The performance of the proposed algorithms is verified in a 75 km 40 GBaud polarization division multiplexed 16 quadrature amplitude modulation (PDM-16QAM) system. The computational complexity of FNT-CDC + FNT-AEQ scheme we proposed is 68% lower compared with FFT-based scheme.

**Fermat number transform**

The FNT is a type of the number-theoretic transform (NTT). The NTT and INTT of length $N$ are defined by the following equations:

$$X(k) = FNT(x(n)) = \sum_{n=0}^{N-1} x(n)\alpha^{nk} \pmod{F} \quad (1)$$

$$x(n) = IFNT(X(k)) = N^{-1}\sum_{k=0}^{N-1} X(k)\alpha^{-nk} \pmod{F} \quad (2)$$

Where $F$ and $\alpha$ are the modulus and the radix of the number-theoretic transform, respectively. $F$ and $\alpha$ need to satisfy the following equations [5]:

$$F = p_1^{r_1} p_2^{r_2} \cdots p_l^{r_l} \quad (3)$$

$$\alpha^N = 1 \pmod{p_i^{r_i}}, \quad i = 1, 2, \cdots, l \quad (4)$$

Where $p_i$ is the prime factor of $F$. Ordinary number-theoretic transforms still require multiplication and modulus operations. In hardware computation, data exists in binary form, where performing bitwise shifts and addition operations are much less complex than multiplication. When we choose the radix value $\alpha$ as a power of 2, multiplication operations in the transformation will be converted into shift operations [6]. Taking it a step further, by selecting the modulus $F$ as a Fermat number of the form $F_t = 2^{2^t} + 1$ $(t = 0,1,3,\cdots)$, all modulo operations in the NTT can be replaced with bitwise shift and addition operations [7]. The transformation that satisfies the above conditions can be called the Fermat number transform.

The FNT can be expressed as a butterfly-like fast calculation structure [8]. Taking an 8-point 2-radix 17-modulus FNT as an example, it can be efficiently calculated using the structure shown in Fig. 1, with only bitwise shift and addition operations throughout the entire process.

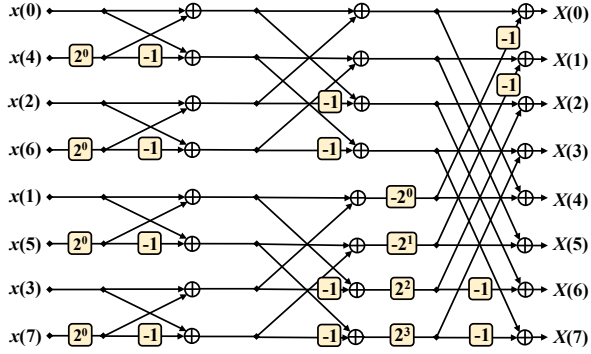

**Fig. 1:** 8-point, 2-radix, 17-modulus FNT butterfly structure.

## Proposed FNT-CDC&AEQ structure

We designed the CDC and AEQ structure based on FNT to meet the DSP requirements of ZR coherent optical communication systems.

FNT operates on data in the finite field of positive integers rather than in the real number field [8]. Therefore, before utilizing FNT to compute convolution, it is necessary to quantize the data. Due to the overflow in the computation process of FNT, the bit width of the transformed sequence is constrained by the modulus [5]. The constraint is represented by the following inequality:

$$|y(m)|_{max} \sum_{n=0}^{N-1} |x(n)| \leq (F_t - 1)/2 \quad (5)$$

Where $x(n)$ and $y(m)$ are two sequences to be convolved, and $F_t$ is the modulus. It should be noted that performing linear convolution via circular convolution requires zero-padding, thus the constraint can be relaxed.

In the CDC, the length of the filter directly determines its ability to compensate for chromatic dispersion [3]. We select the modulus $F_t$ to be 65537 and the radix $\alpha$ to be $\sqrt{2}$, with a transform sequence length $N$ of 64. The signal has a bit width of 5 and the CDC-taps have a bit width of 6.

In the AEQ, bit width limitation makes it difficult for a single FNT to meet the update requirements of taps. We implement a radially directed equalizer (RDE) to update the taps of the filter. The AEQ structure is 4x4 RV-MIMO [9]. The taps update is based on the following equation:

$$w'_{S_1-S_2} = w_{S_1-S_2} + \mu e_{S_2} Sig_{in\_S_1} Sig_{out\_S_2} \quad (6)$$

Where $S_1$ and $S_2$ take *XI, XQ, YI, YQ*. $w_{S_1-S_2}$ represents the original filter tap, $w'_{S_1-S_2}$ represents the updated filter tap, $e_{S_2}$ denotes the error function, $Sig_{in\_S_1}$ and $Sig_{out\_S_2}$ are the input and output signal sequences, $\mu$ is the update step size. The taps update amount $\Delta w = \mu e_{S_2} Sig_{in\_S_1} Sig_{out\_S_2}$ for one step is usually on the order of $10^{-4}$. Therefore, the bit width of the filter taps needs to exceed 14. We divide the filter taps into two groups based on bit width, consisting of the first 7 bits and the last 7 bits of the data, and perform convolution on each group separately. The final result is obtained by the weighted sum of two convolution results. Corresponding to the bit width of each group of filter taps, we choose the modulus $F_t$ to be 65537 and the radix $\alpha$ to be 2, with a transform sequence length $N$ of 32, while the bit width of the signal is 5.

Additionally, when calculating complex convolution using the Fermat number transform, complex multiplication can be decomposed into two real multiplications [10], as shown in equation (7) and (8).

$$z_n = -2^{q-1}(IFNT\{(X_k + 2^{q/2}\hat{X}_k)(Y_k + 2^{q/2}\hat{Y}_k) \\ + (X_k - 2^{q/2}\hat{X}_k)(Y_k - 2^{q/2}\hat{Y}_k)\}) \quad (7)$$

$$\hat{z}_n = -2^{q/2-1}(IFNT\{(X_k + 2^{q/2}\hat{X}_k)(Y_k + 2^{q/2}\hat{Y}_k) \\ - (X_k - 2^{q/2}\hat{X}_k)(Y_k - 2^{q/2}\hat{Y}_k)\}) \quad (8)$$

Where $q$ is $2^t$, $X_k$, $Y_k$ are the real parts, and $\hat{X}_k$, $\hat{Y}_k$ are the imaginary parts of $x_n$, $y_n$ in Fermat domain. $z_n$ and $\hat{z}_n$ are the real and the imaginary parts of the convolution result, respectively. We apply this method to the FNT-CDC structure to further reduce its complexity. The FNT-CDC and FNT-AEQ structure are illustrated in Fig. 2 and 3.

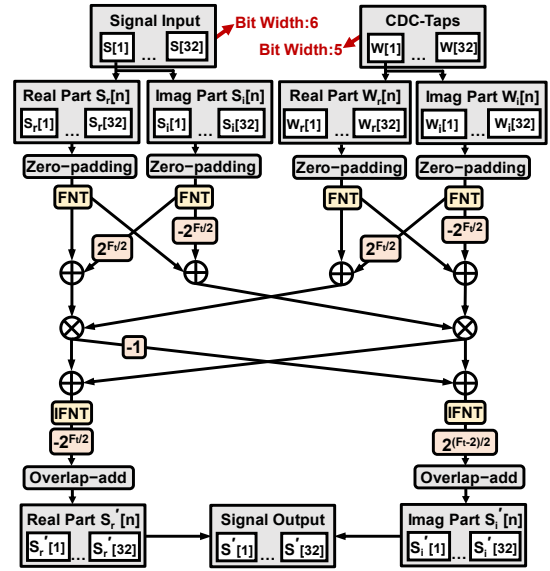

**Fig. 2:** FNT-CDC structure.

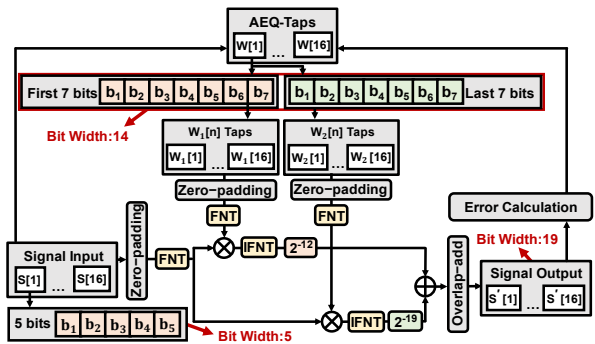

**Fig. 3:** FNT-AEQ structure.

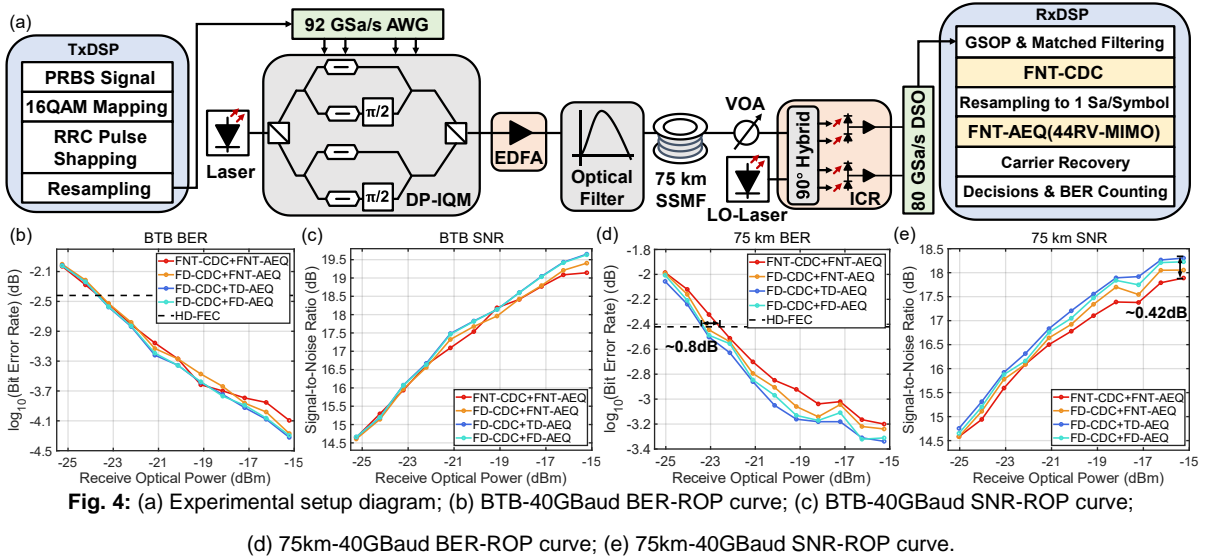

**Fig. 4:** (a) Experimental setup diagram; (b) BTB-40GBaud BER-ROP curve; (c) BTB-40GBaud SNR-ROP curve; (d) 75km-40GBaud BER-ROP curve; (e) 75km-40GBaud SNR-ROP curve.

For the implementation of a DSP algorithm on an ASIC or an FPGA, the effort for a multiplier is much higher than for an adder [3]. We compare the complexity of algorithms by counting the number of real multiplications required for calculating one symbol. TD-AEQ utilizes linear convolution to compute output sequences, while FNT-AEQ and FD-AEQ schemes yield cyclic convolution. To ensure equivalence between frequency domain and time domain convolution, when comparing the complexity and performance of TD-CDC/TD-AEQ with other schemes, the length of the filter taps is set to half that of the other schemes. The Tab. 1 shows the number of real multiplications per symbol varying with length $N$ in different schemes. In the context of our experiment, we compare the proposed 64-point FNT-CDC + 32-point FNT-AEQ scheme with the 64-point FD-CDC + 32-point FD-AEQ and 32-point TD-CDC + 16-point TD-AEQ schemes. The number of real multiplications of FNT-CDC and FNT-AEQ is reduced by 89% and 58% compared with FD-CDC and FD-AEQ, respectively. Taking both into consideration, we can conclude that the FNT-CDC + FNT-AEQ scheme reduces the computational complexity by 68% compared to conventional FFT-based scheme.

**Tab. 1:** The number of real multiplications in different schemes

| Scheme | Real Multiplications |
| --- | --- |
| FNT-CDC | 8 |
| FD-CDC | $12\log_2 N+12$ |
| TD-CDC | $6N$ |
| FNT-AEQ | 64 |
| FD-AEQ | $24\log_2 N+32$ |
| TD-AEQ | $16N$ |

**Experimental verification**

The experimental system is illustrated in Fig. 4(a). At the Tx, we use a root-raised cosine (RRC) roll-off pulse shaping to generate a 40 GBaud DP-16QAM signal. The signal is amplified by an EDFA, and an optical filter is used to remove out-of-band noise. Then, the signal is transmitted through 75 km of standard single mode fibres (SSMFs). At the Rx, a variable optical attenuator (VOA) is used to adjust the received optical power (ROP), and the digital storage oscilloscope (DSO) samples at 80 GHz. After GSOP and matched filtering, the proposed FNT-CDC and FNT-AEQ are used, followed by carrier recovery. The signal is resampled to 1 Sa/symbol before entering AEQ block.

In Fig. 4(b), (c), (d), and (e), we present the BER-ROP/SNR-ROP curves for different schemes in BTB and 75 km scenarios to evaluate the performance of each scheme. We can find from Fig. 4(b) and (c) that our proposed FNT-CDC + FNT-AEQ scheme has negligible performance penalty compared with conventional floating-point-FD-CDC+TD-AEQ scheme in BTB scenario. In ZR transmission, the FNT-CDC + FNT-AEQ scheme has a receiver sensitivity penalty of only 0.8 dB compared with floating-point-FD-CDC+TD-AEQ scheme at hard decision forward error correction (HD-FEC) limit, as shown in Fig. 4(d). The maximum SNR penalty is 0.42 dB, as indicated in Fig. 4(e).

**Conclusions**

We demonstrated the Fermat number transform based CDC and AEQ in coherent data centre interconnection, as verified through a 75-km 40-Gbaud DP-16QAM transmission in experiments. The results indicate that the proposed FNT-SDC+FNT-AEQ scheme, compared with FFT-based floating-point-SDC+AEQ, reduces computational complexity by 68% while ensuring less than 0.8 dB receiver sensitivity penalty.

**Acknowledgements**

This work was supported by the National Natural Science Foundation of China under Grants 62225110, 61931010 and the Major Program (JD) of Hubei Province (2023BAA001-1).